\documentclass[letter]{aa}     

\usepackage{graphicx}
\usepackage{txfonts}
\usepackage{lipsum}
\usepackage{subcaption}         
\usepackage{lscape}             
\usepackage{placeins}           
\usepackage{verbatim}
\usepackage{tikz}
\usetikzlibrary{arrows.meta, positioning, quotes}

\usepackage[colorlinks=true,citecolor=blue]{hyperref}

\usepackage{multirow}
\usepackage{threeparttable}
\usepackage{booktabs} 

\usepackage[normalem]{ulem}  
\newcommand\redsout{\bgroup\markoverwith{\textcolor{red}{\rule[0.5ex]{10pt}{2.pt}}}\ULon}  
\usepackage{xcolor}

\begin{document}

\title{Tango of Titans: Centaurus~A and M\,83 as a Local Group analog}



\author{David Benisty\inst{1}    
          \and Noam Libeskind \inst{1} \and Dmitry Makarov \inst{2} }

\institute{$^{1}\,$ Leibniz-Institut f\"ur Astrophysik Potsdam (AIP), An der Sternwarte 16, 14482 Potsdam, Germany\\
 \email{benidav@aip.de, nlibeskind@aip.de}\\
$^{2}\,$ Special Astrophysical Observatory, Russian Academy of Sciences, Nizhnii Arkhyz, 369167 Russia\\
  \email{dim@sao.ru}}

   \date{}

 
  \abstract
{Centaurus A (Cen~A) and M~83 form one of the most massive galaxy pairs in the nearby Universe. Although their observed heliocentric velocities suggest motion that is not obviously indicative of mutual attraction, this work presents evidence that Cen~A and M~83 are in fact gravitationally attracted toward each other, exhibiting a dynamical interaction analogous to the binary-like motion of the Milky Way and Andromeda in the Local Group (LG). Using the timing argument, calibrated with analog galaxy pairs from the AbacusSummit simulation, we estimated the total mass of the Cen A/M83 system under the assumption that the line-of-sight velocity is dominated by motion toward the system's barycenter. This yields a total mass of $(6.36 \pm  1.30) \cdot 10^{12}\, M_\odot$. The inferred mass agrees well with independent estimates based on virial mass measurements and $K$-band luminosity-to-mass ratios. Together, the consistent bound signature and robust mass determination highlight the Cen A/M83 system as a compelling nearby analog to the LG. A further discussion of NGC 4945 as a main perturber (similar to the Large Magellanic Cloud for the LG) of Cen A is also presented.}

\keywords{galaxy pairs and groups --
Timing Argument --
mass estimation --
Cen\,A/M\,83 --
dark matter halos --
$\Lambda$CDM cosmology
}

   \maketitle
\nolinenumbers

\section{Introduction}
The halo mass is an important quantity with a few independent methods, that requires assumptions about the dynamical state of the system. For example, estimates of virial mass assume that the galaxy or group is in dynamical equilibrium \citep{Bahcall:1981}, while methods based on spherical symmetry or rotation curves rely on idealized geometric or kinematic assumptions \citep{mo2010galaxy,Courteau:2013cjm}. A unique case occurs in galaxy pairs that are sufficiently close to interact gravitationally but not so close as to be violently merging. In these systems, the mutual motion of the galaxies can provide a direct probe of the combined mass. The classical example is the Milky Way--Andromeda (M31) pair in the Local Group (LG), where the present separation and relative velocity allow the total mass to be inferred using the Timing Argument (TA; \citealt{Li:2007eg,Penarrubia:2014oda,Wang:2019ubx,Hartl:2021aio,Benisty:2022ive,Benisty:2024lsz,Strigari:2025nqa}). The TA models the two galaxies as an isolated two-body system that initially followed the Hubble expansion and is now collapsing under mutual gravity. By combining the current separation, the age of the Universe, and the measured radial velocity, one can estimate the total mass of the pair~\citep{vanderMarel:2007yw,Partridge:2013dsa,McLeod:2016bjk,Zhao:2013uya,Banik:2018zhm,Penarrubia:2015hqa,Chamberlain:2022fqr,Benisty:2020kys}.

Here we extend this reasoning to the nearby Centaurus A (Cen~A)--M\,83 system, which has long been assumed to be simply receding with the cosmic expansion but is instead a compelling dynamical analog to the LG. The Cen A/M83 complex contains two dominant subgroups: the massive elliptical Cen~A (NGC~5128; the virial mass estimated from the dispersion of its satellite is $\sim 6\cdot10^{12}\,M_\odot$) and the star-forming spiral M\,83 (NGC~5236; with a virial mass of $\sim[1.3$--$3]\cdot10^{12}\,M_\odot$; \citealt{Karachentsev:2006,Karachentsev:2007AJ,Muller:2015,Muller:2016,Muller:2019,Muller:2021c,Faucher:2025blj}). 

\section{Two-body motion on the sky}
\label{sec:TA}

\subsection{Initial conditions}
The space-time governing the dynamics of a two-body system in an expanding universe can be effectively modeled by considering a test particle with reduced mass moving under the influence of a central gravitational potential, embedded within a cosmological background~\citep{Sereno:2007tt, Faraoni:2007es, Nandra:2011ug}. This framework captures the combined effects of local gravitational attraction and large-scale cosmic expansion in $\Lambda$ Cold Dark Matter ($\Lambda$CDM). Cosmological expansion impact the evolution of gravitationally bound systems, such as galaxy pairs and binary halos. In the weak-field, the equation of motion for the test particle becomes \citep{Banik:2015nia,Banik:2016emv,Benisty:2024lsz}
\begin{equation}
\ddot{r} = -\frac{G M}{r^2} + \frac{l^2}{r^3} + \frac{\ddot{a}}{a} r,
\label{eq:eomDaDt}
\end{equation}
where $r$ is the separation between the two bodies, $M$ is the total mass of the system, $G$ is Newton's gravitational constant, and $l = r_0 \cdot v_{\tan}$ is the specific angular momentum (i.e., per the reduced mass), with $r_0$ and $v_{\tan}$ denoting the current physical separation and tangential velocity, respectively. The final term, $\ddot{a}/a \cdot r$, encapsulates the effect of cosmic acceleration, derived from the evolution of the scale factor for $\Lambda$CDM using \textit{Planck} data~\citep{Planck:2018vyg}, where the cosmic acceleration is given by $\ddot{a}/a = H_0^2 (\Omega_{\Lambda,0} - \Omega_{m,0}/2a^3)$. This term approaches $H_0^2 \Omega_{\Lambda,0} = \Lambda c^2/3$ asymptotically. To constrain the system's mass, we numerically integrated the equation of motion backward in time, starting from the present-day separation and radial velocity. The integration was subjected to the boundary condition that as $t \to 0$, the galaxies emerged from initial Hubble flow: $v_{\text{pec}} \equiv v_{\mathrm{rad}} - H(t) \cdot r \rightarrow 0$.

\begin{figure}[t!]
    \centering
    \includegraphics[width=0.75\linewidth]{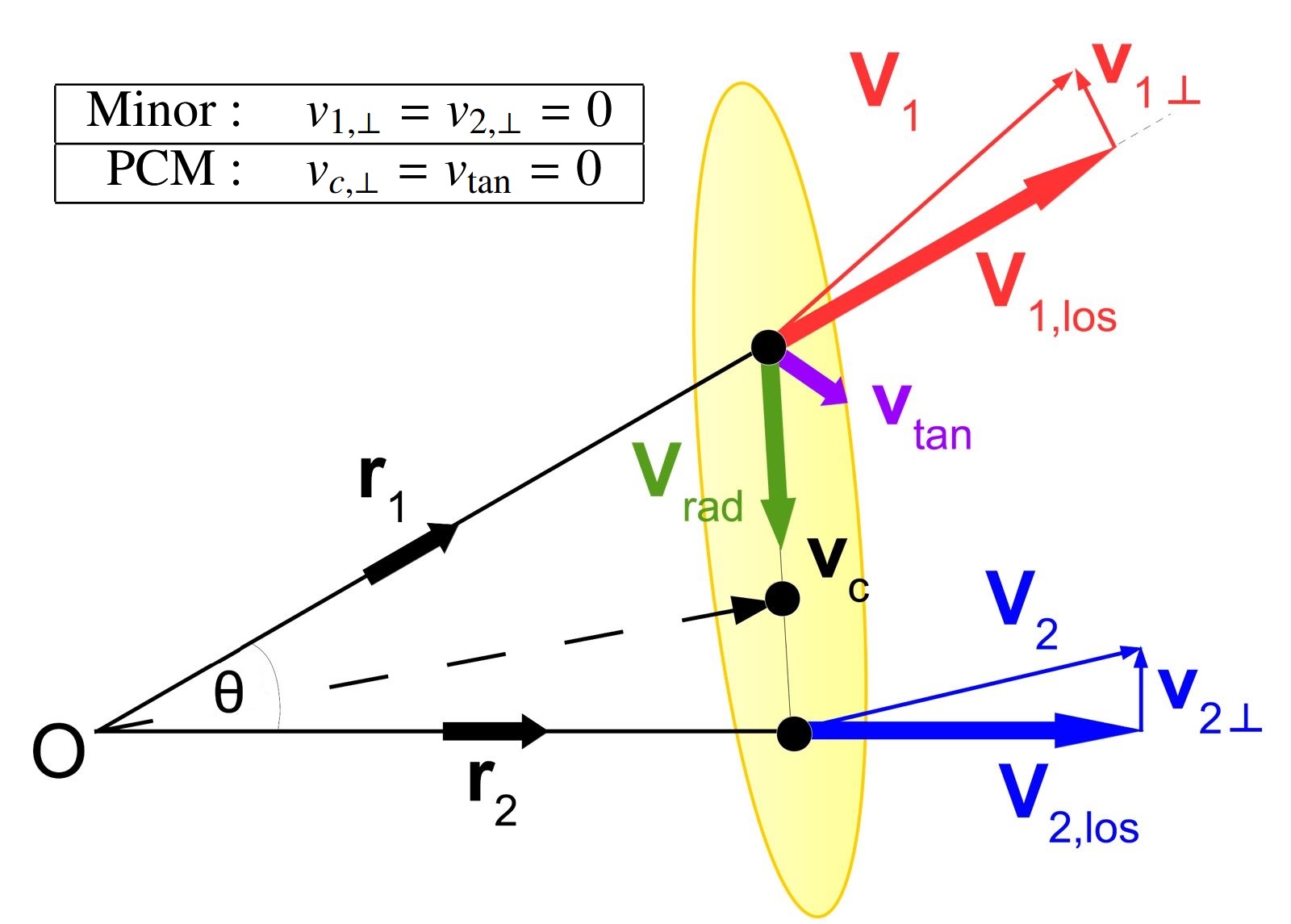}
\caption{Relative motion of two galaxies, labeled galaxy 1 and galaxy 2. The vector $v_1$ ($v_2$) represents the velocity of galaxy 1 (galaxy 2). The angle $\theta$ is measured between $r_1$ and the line connecting the two galaxies. $v_c$ is the LoS velocity of the CoM of the pair. The assumptions for different infall models are shown in the top left. }
    \label{fig:rel_motion}
\end{figure}

\begin{table}
\caption{Data used in this study. }
\label{tab:galaxy_data}

{\small
\centering
\begin{threeparttable}
\setlength{\tabcolsep}{3pt}
\begin{tabular}{|l|cc|c|cc|c|}
\hline\hline
 &
RA&
Dec&
$D^\dagger$&
$v_\mathrm{hel}$& 
$v_\mathrm{LG}$& 
$L_{K}^\ddagger$\\[2pt]
\cline{2-3}\cline{5-6}
&
\multicolumn{2}{c}{$^\circ$} &
Mpc &
\multicolumn{2}{c}{km\,s$^{-1}$} & 
$\times 10^{10}$~$L_\sun$ \\
\hline
Cen\,A  & 201.37 & $-43.02$ & $3.68 \pm 0.05$ & $547 \pm 5 $ \tnote{a} & $ 311.5 \pm 5.8 $ & 7.76 \\ 
M\,83   & 204.25 & $-29.87$ & $4.90 \pm 0.08$ & $518 \pm 3 $ \tnote{b} & $ 316.2 \pm 4.2 $ & 7.24 \\ 
\hline\hline
\end{tabular}

\begin{tablenotes}
\item[$\dagger$] \citet{2021AJ....162...80A}\, $^{\ddagger}$ \citet{Karachentsev:2021vau}
\item[a] \citet{1992ApJS...83...29S}\, $^{b}$ \citet{2008AJ....136.2563W}
\end{tablenotes}

\end{threeparttable}
}

\end{table}

\subsection{LG-like systems observed from the outside}
\label{sec:LG-like}

To determine the TA mass, we needed the radial velocity between the galaxies. Figure~\ref{fig:rel_motion} illustrates a pair of galaxies observed by an external observer with LoS velocities $v_1$ and $v_2$. The motion of galaxy 1 as seen by galaxy 2 is decomposed into its the radial component ($\vec{v}_{\mathrm{rad}}$) and the tangential one ($\vec{v}_{\tan}$). Any velocity can be projected onto the LoS component ($v_{\text{los}}$) and the perpendicular component on the plane between the observer and the two galaxies ($v_{2,\perp}$); see Appendix~\ref{sec:math} for the full geometrical derivation. The $v_{2,\perp}$ and $v_{1,\perp}$ are unknown, but one can assume that at least one is zero, which enables two different realizations~\citep{Karachentsev:2006,Wagner:2025wrp}. 

One assumption is that the perpendicular velocities are low, $v_{2,\perp} =  v_{1,\perp} = 0$, which is often called the minor infall model. This is essentially a minimal model where only the actual observed LoS velocities are considered and the other components are assumed to be negligible. In this case, $v_{\mathrm{rad}}$ can be written as
\begin{equation}
v_{\mathrm{rad}} = \ v_{1,\text{los}} \bar{r}_1 + v_{2,\text{los}} \bar{r}_2 - \cos \theta \left( v_{1,\text{los}} \bar{r}_2 + v_{2,\text{los}} \bar{r}_1 \right),
\label{eq:minor_infall}
\end{equation}
where $\bar{r}_{i} \equiv r_{i}/r_{12}$ is the ratio between the distance to the galaxy ($r_{i}$) and the 3D distance between the galaxies ($r_{12}$). 

Another assumption is that the barycenter of the galaxy pair has only a LoS component, i.e., $v_{c,\perp} = 0$. Using the mass ratio of the galaxies, which determines the center of mass (CoM) on the sky, we get the projection CoM (PCM) infall model:
\begin{equation}
 v_\mathrm{rad} = \frac{\bar{m}_1 \bar{r}_1 v_{2,\text{los}} -\bar{m}_2 \bar{r}_2 v_{1,\text{los}} - \cos\theta \left( \bar{m}_1 \bar{r}_1 v_{1,\text{los}} - \bar{m}_2 \bar{r}_2 v_{2,\text{los}}\right)}{ \cos\theta \left(\bar{m}_1 \bar{r}_1^2-\bar{m}_2 \bar{r}_2 ^2\right) + \bar{r}_1 \bar{r}_2 \Delta \bar{m}},
\label{eq:radial_com_major}
\end{equation}
where $\Delta \bar{m} = \bar{m}_1 - \bar{m}_2$. The advantage of this projection is the lack of the dependence of the $v_{\perp}$.

The PCM model uses the unknown ratio of the two masses, and applying the model requires employing a prior assumption on the probability of that ratio. These assumptions are oversimplified since the dynamics are likely more complicated. However, since tangential motions are impossible to measure, we made these assumptions for the sake of simplicity.

\begin{figure}[t!]
\centering
\includegraphics[width=0.45\textwidth]{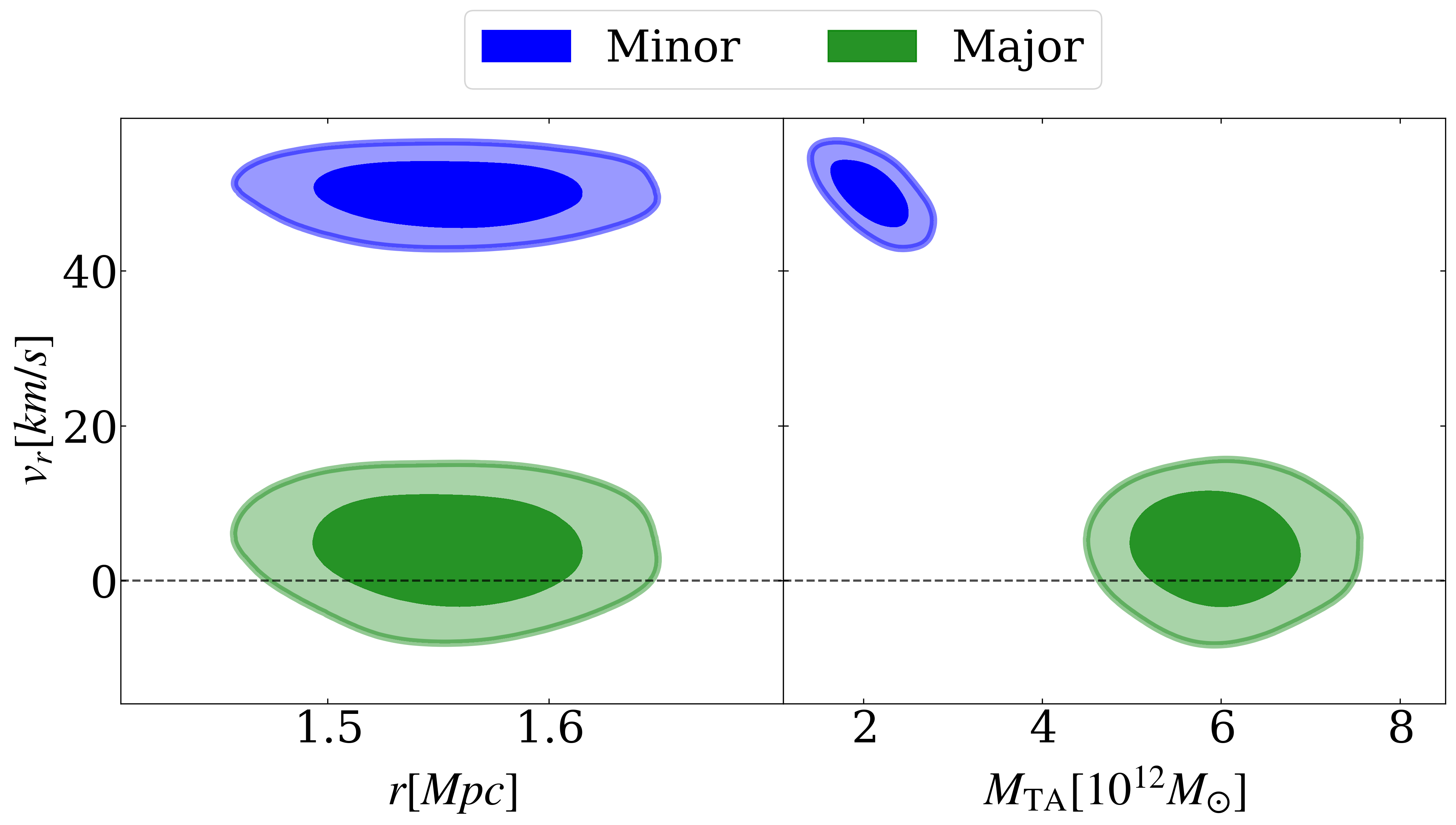}
\caption{Inferred quantities for the Cen A--M83 radial motion. }
\label{fig:data}
\end{figure}

\begin{table}[t!]
\caption{The values for fig~\ref{fig:data} }
    \centering
\begin{tabular}{|c|c|c|c|}
\hline
Model &
$r\,[\mathrm{Mpc}]$ &
$v_{\mathrm{rad}}$~km\,s$^{-1}$&
$M_{\text{TA}}\,[10^{12} M_{\odot}]$ \\
\hline\hline
Minor & \multirow{2}{*}{$1.55 \pm 0.08$} & $+50 \pm 3$ & $2.1 \pm 0.5$ \\
\cline{1-1}\cline{3-4}
PCM   &                  & $-6 \pm 10$ & $6.0 \pm 1.2$ \\
\hline\hline
\end{tabular}
\label{tab:posterior}
\end{table}

\section{Mass determination}
\label{sec:mass}

\subsection{Data and relative motion}

\begin{figure*}[t!]
\centering
\includegraphics[width=0.95\textwidth]{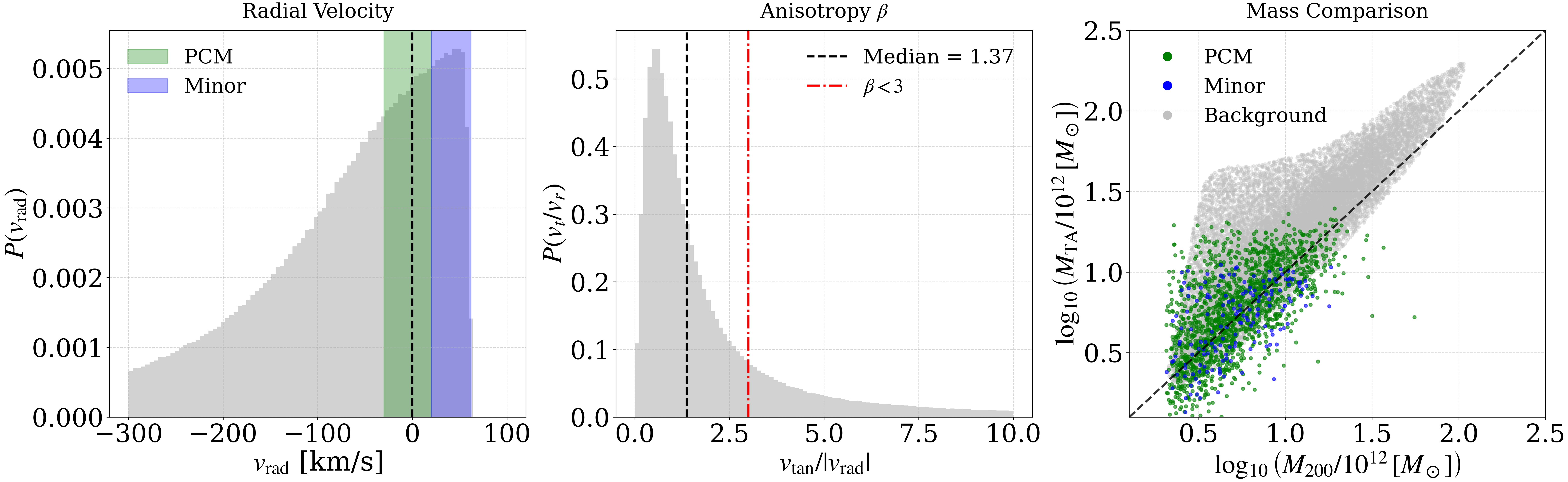}
\caption{Kinematic properties of the simulated pair sample. The panels show the radial velocity cuts used to define the PCM (green) and minor (blue) candidates (left); the velocity anisotropy distribution (middle); and the resulting $M_{TA}$ vs. $M_{200}$ relation for each group (right).}
\label{fig:cor_CenAM83}
\end{figure*}

Both galaxies, Cen\,A and M\,83, have a large number of distance measurements derived via various methods.
We chose to use homogeneous data from the color-magnitude diagrams and the Tip of the Red Giant Branch Distance Catalog~\citep{2021AJ....162...80A} of the Extragalactic Distance Database~\citep[EDD;][]{2009AJ....138..323T}, which are based on the old stellar population as standard candles.
The high-precision distance moduli of $\mu \equiv (m-M)_0 = 27.83^{+0.03}_{-0.04}\mathrm{(stat)}\pm0.04\mathrm{(sys)}$ and $\mu = 28.45^{+0.04}_{-0.03}\mathrm{(stat)}\pm0.04\mathrm{(sys)}$ for Cen\,A and M\,83, respectively, were obtained using \textit{Hubble} Space Telescope observations with the Advanced Camera for Surveys. As the best heliocentric velocities, we used the optical measurement of $v_\mathrm{hel} = 547\pm5$ by \citet{1992ApJS...83...29S} for Cen\,A, which perfectly matches the radio data in EDD, $547\pm20$~\citep{2009AJ....138.1938C}; and $v_\mathrm{hel} = 518.0\pm2.6$ for M\,83, which was measured with high spectral and spatial resolution as part of the HI Nearby Galaxy Survey~\citep[THINGS;][]{2008AJ....136.2563W}.
Heliocentric velocities were converted to the LG barycenter using the solar apex vector~\citep{2025A&A...698A.178M}, determined from an analysis of the Hubble flow in the vicinity of the LG.
Data for Cen\,A and M\,83 are systematized in Table~\ref{tab:galaxy_data}.

Both mass models were applied to the distances and LoS velocities in the LG barycenter reference frame of the two dominant galaxies in the Cen\,A/M\,83 system.
The TA mass of the system was assessed by performing Monte Carlo realizations of the uncertainties on the distances and the LoS velocities as provided in the database. The errors on the distance moduli were assumed to be normally distributed. The application of the PCM model requires an assumption on the unknown mass ratio, which we adopted from the $K$-band luminosity of the two galaxies~\citep{Karachentsev:2021vau}.

Figure~\ref{fig:data} illustrates the distributions of inferred distances and infall velocities derived from the data in Table~\ref{tab:galaxy_data}. Both models robustly constrain the radial component of the relative velocity, $v_{\rm rad}$, to a notably small magnitude. The PCM suggests that Cen\,A and M\,83 could have a negative radial velocity. Such a low-velocity regime is consistent with scenarios in which the galaxies are loosely bound or evolving quasi-independently. While \cite{Karachentsev:2006ww} argued for a positive radial velocity as indicated by the minor infall, our application of the PCM assumption predicts a possible negative radial velocity, providing the first evidence that this system is gravitationally bound and infalling.

\subsection{TA mass and simulation inference}

The various versions of the TA model have been tested with the \textit{AbacusSummit} simulation~\citep{Maksimova:2021ynf}. 
This fits our task particularly well since it includes a high-resolution run with particle masses of $m_{\rm part} = 2 \times 10^9 \, h^{-1}M_{\odot}$ within a large simulation box of 2~Gpc\,$h^{-1}$ size for multiple cosmological models.
Here we used the data for the fiducial \textit{Planck} 2018 cosmology~\citep{Planck:2018vyg}. 
The resulting $z=0$ halo catalog contains approximately $12 \times 10^6$ halos with masses above $10^{11}\,M_{\odot}$, identified via a spherical overdensity finder applied after a friends-of-friends grouping~\citep{Maksimova:2021ynf}. The pair-finding algorithm selects halos with masses between $5\times10^{11}$ and $5\times10^{13}\,M_{\odot}$ and associates each with a single, more massive neighbor. A halo qualifies as a valid partner if it lies within 2 Mpc in 3D separation and if no third halo with $M_{\rm halo}\geq10^{11}M_{\odot}$ is located closer than the chosen companion. To avoid duplication, each halo was allowed to be part of only one pair. Additionally, the relative radial velocity between the two halos must not exceed the Hubble-flow velocity (about 100~km\,s$^{-1}$) for the CenA/M83 pair.

To model the velocity components, we used the radial velocity using both the minor and PCM infall assumptions. The tangential velocity ($v_{\tan}$) was sampled via the distribution of $\beta = v_{\tan} / |v_\mathrm{rad}|$ measured from the cosmological analogs. We tested the TA against the analog pairs under the ideal assumption of exact knowledge of separation and relative velocity, as in~\citet{Li:2007eg}. Solving Eq.~\ref{eq:eomDaDt} for the separation and with the radial velocity giving the TA predicted mass, the minor infall model yields a low TA mass of $(2.1 \pm 0.5)\cdot 10^{12} \,M_{\odot }$ and the PCM a higher TA mass of $(6.0 \pm 1.2)\cdot 10^{12} \,M_{\odot }$. 

\begin{table}[t!]
    \centering
\caption{ {Ratio of the simulation-based mass to the TA mass (left) vs. the final TA + simulation inference mass (right).}}
\begin{tabular}{|c|c|c|}
\hline\hline
System   & $M_{\text{200}} / M_{\text{TA}}$ &  $M_{\text{TA + Sim}}$ \\      
\hline\hline
Cen\,A--M\,83 [Minor]  & $1.21 \pm 0.65$ & $1.86 \pm 0.97$ \\       
\hline
Cen\,A--M\,83 [PCM] & $ 0.80 \pm 0.49$ & $6.36 \pm 1.31$  \\
\hline\hline 
\end{tabular}
\label{tab:mass_rat}
\end{table}

Simulation-informed correction ($M_{\text{200}}/M_{\text{TA}}$) was applied to account for the systematics inherent to the TA method. Figure~\ref{fig:cor_CenAM83} shows the corresponding correlation between TA masses and the virial masses of the simulation-based analogs. The minor infall solution, representing an unbound or loosely bound configuration, yields a mass of $(1.21 \pm 0.65) \times 10^{12}\,M_{\odot}$, whereas the PCM model predicts a significantly higher value of $\left(6.36 \pm 1.31\right)  \times 10^{12}\,M_{\odot}$. 

There are cases where the TA crosses the 1:1 correlation~\citep{Li:2007eg} and in other cases introduces some bias~\citep{Hartl:2021aio,Benisty:2022ive}, which depends on the phase space selection (the radial the tangential velocity ranges). However, in our case statistics allowed for the 1:1 correlation when the radial velocity was much lower then the Hubble flow (PCM), and the minor infall introduces larger biases (see Table~\ref{tab:mass_rat}).

The K-band luminosity-to-mass relation, i.e., Eq.~8 from~\citet{Tully:2017}, provides an independent constraint. Adopting the 2MASS $K$-band luminosities from Table~\ref{tab:galaxy_data}, we obtained a combined mass of $M_{\rm total} = (8.26 \pm 1.18)\cdot 10^{12} M_\odot$. Virial analyses of the Cen\,A subgroup and the M\,83 subgroup ~\citep{Karachentsev:2006ww}
yield a combined system mass of $(6.0 \pm 1.4)\cdot 10^{12}M_\odot$. Multi-Unit Spectroscopic Explorer satellite observations yielded a M\,83 subgroup total mass of $8.9^{+1.89}_{-1.94}\cdot 10^{12}M_\odot$~\citep{Muller:2025}, and \citet{Faucher:2025blj} find $\left(7 \pm 2 \right) \cdot 10^{12}M_\odot$, which is consistent with our PCM mass estimates. The Hubble-flow fit yields lower masses of about $\sim 2\cdot10^{12}M_\odot$~\citep{Peirani:2008qs,DelPopolo:2022sev,Faucher:2025blj}, similar to our minor infall estimator. As~\citet{Faucher:2025blj} show, the mass based on the Hubble flow method is biased low since the M \,83 galaxy is located too close to the turnaround of the system, and in such a case the Hubble flow fit model breaks.

\section{Discussion}
\label{sec:dis}

\begin{figure}[t!]
\centering
\includegraphics[width=0.45\textwidth]{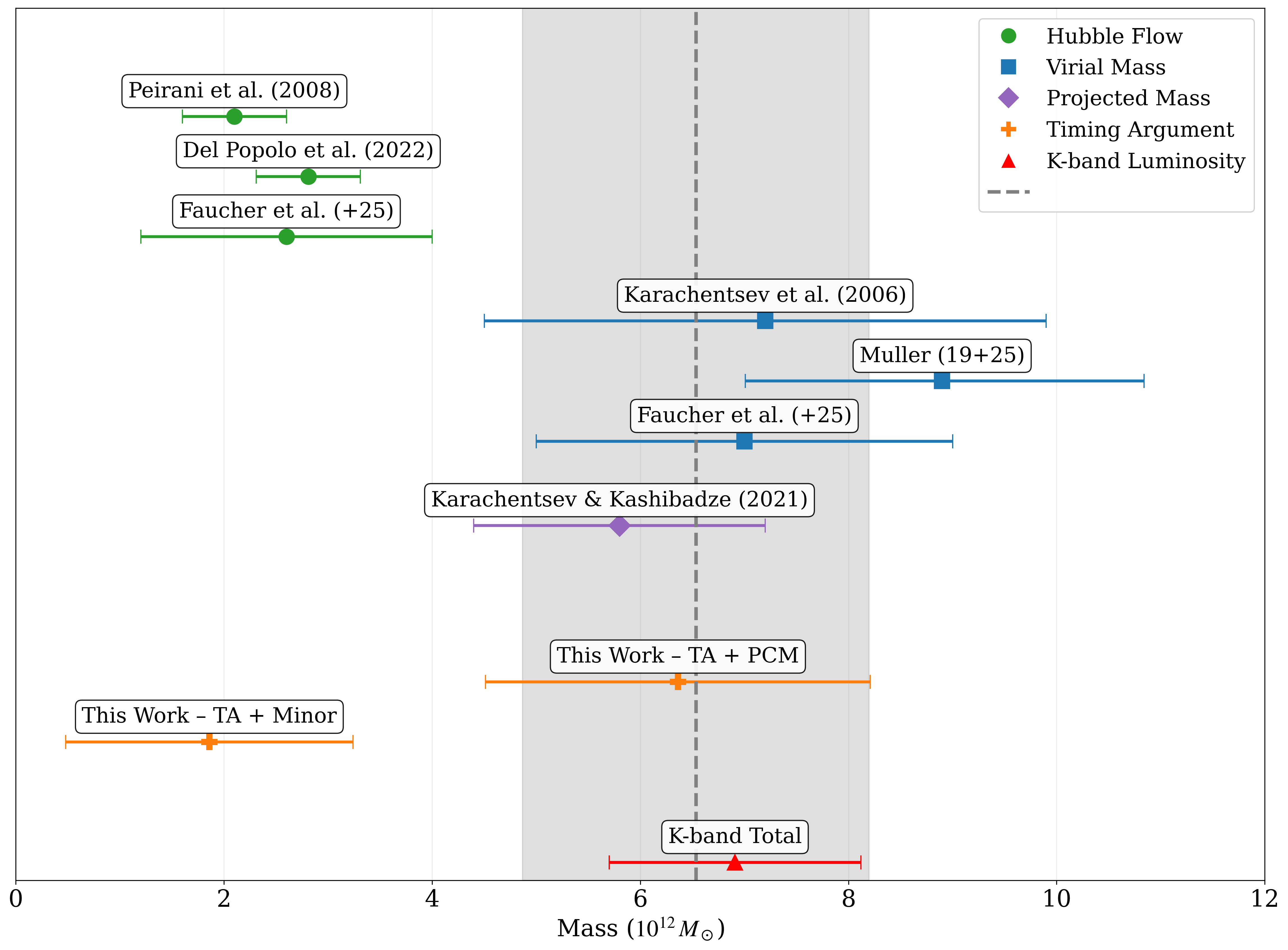}
\caption{Total mass estimates for the Cen\,A--M\,83 galaxy system from the virial, projected mass, Hubble-flow, and TA methods. The PCM solution agrees well with dynamical and stellar-based estimates, while the minor solution aligns more closely with Hubble-flow prediction.}
\label{fig:lit_mass}
\end{figure}

This work presents the first application of the TA to the Cen A--M83 system, providing an independent dynamical mass constraint for this nearby galaxy pair. The PCM relative motion predicts a mass of $M_{\text{TA,major}} = (6.36 \pm 1.31) \cdot 10^{12} \, M_\odot$, with an infall velocity of $v_{\mathrm{rad}}^{\rm Cen\,A-M\,83} = -6 \pm 10$~km\,s$^{-1}$. When placed in the broader context of other mass estimators (Fig.~\ref{fig:lit_mass}), the PCM mass is fully consistent with both virial mass determinations and K-band luminosity-based estimates, which cluster around $[6$--$9]\cdot10^{12} M_\odot$.

Backward integration of the TA equations, using a K-band luminosity mass prior, yields a present-day radial velocity ($v_{\mathrm{rad}}$) of $-31 \pm 51$~km\,s$^{-1}$, which favors the PCM scenario. Both TA solutions imply bound or marginally bound orbits. At the observed separation, the Hubble-flow velocity exceeds 100~km\,s$^{-1}$, which is significantly higher than either the minor or PCM radial velocity predictions. The corresponding specific orbital energies for the two TA configurations are negative, $\epsilon \sim -0.5 \cdot 10^3$~km$^2$s$^{-2}$ (minor) and $\epsilon \sim -2 \cdot 10^3$~km$^2$s$^{-2}$ (PCM), confirming that both lie in the bound regime, where $\epsilon$ is the total energy per the reduced mass.

The turnaround of the group further reinforces this conclusion~\citep{bib:Sandage1986}. The Cen A turnaround radius is estimated to be $\sim 1.4$~Mpc~\citep{Karachentsev:2007AJ}, only slightly smaller than the current separation of $\sim 1.5$~Mpc. Relative to Cen A alone, M83 may reside just outside the collapsing region, which would allow for a small positive radial velocity, as predicted by the minor infall model. However, if we consider the combined gravitational potential of the full CenA/M83 system, the turnaround radius is larger. Relative to the barycenter, both galaxies lie within the collapsing region, consistent with the PCM prediction of a modest approach speed.

We also address the potential influence of NGC\,4945. Although substantial ($L_K \approx 0.6 L_{K,{\rm Cen\,A}}$), its dynamical mass is significantly lower ($\sim 20\%$ of Cen\,A). Located $0.5$\,Mpc away and on the opposite side of the sky from M83, it forms a tight, bound subsystem with Cen\,A similar to the LMC--Milky Way pair, which will reduce the predicted mass~\citep{Penarrubia:2015hqa,Benisty:2022ive}. We therefore consider it to be dynamically associated with the Cen\,A complex rather than a distinct third body; thus, our derived timing mass encompasses the extended Cen\,A/NGC\,4945 system.

\begin{acknowledgements}
We thank Marcel Pawlowski, Jenny Wagner, Yehuda Hoffman and Igor D. Karachentsev for useful discussions. DB is supported by a Minerva Fellowship of the Minerva Stiftung Gesellschaft fuer die Forschung mbH and acknowledges the contribution of the COST Actions CA23130 and CA21136. NIL acknowledges funding from the European Union Horizon Europe research and innovation program (EXCOSM, grant No.101159513). DIM is supported by the Russian Science Foundation grant \textnumero~24--12--00277.
\end{acknowledgements}

%
\bibliographystyle{aa} 
\bibliography{ref.bib} 

\appendix

\section{Mathematical derivation}
\label{sec:math}
To compute the physical separation ($r_{\rm{12}}$) between the two galaxies given their angular separation ($\theta$), we used

\begin{equation}
r_{\rm{12}}^2 = r_{1}^2 + r_{2}^2 - 2 r_{1} r_{2} \cos\theta,
\end{equation}where $r_1$ and $r_2$ are the distances of the galaxies from the observer. The geometry and velocity components are illustrated in Fig.~\ref{fig:rel_motion}, showing two galaxies with LoS velocities $v_1$ and $v_2$. The motion of galaxy 2 as seen by galaxy 1 is decomposed into a radial component ($\vec{v}_{\mathrm{rad}}$) and a tangential component ($\vec{v}_{\tan}$) such that the full relative velocity is

\begin{equation}
\vec{v}_1 - \vec{v}_2 \equiv \vec{v}_{\mathrm{rad}} + \vec{v}_{\tan}. 
\label{eq:rel_mot}
\end{equation}

Each velocity vector can be written as a combination of three directions: the LoS ($\hat{r}^{\text{los}}_i$), the direction perpendicular to the observer-galaxy plane ($\hat{r}_{i,\perp}$), and the direction perpendicular to the triangle formed by the observer and the galaxy pair ($\hat{r}^{\text{out}}_{i,\perp}$). For each galaxy $i = 1, 2$:

\begin{equation}
\vec{v}_{i} = v_{i,\text{los}} \hat{r}_i + v_{i,\perp} \hat{r}_{i,\perp} + v^{\text{out}}_{i,\perp} \hat{r}^{\text{out}}_{i,\perp}.
\end{equation}

Projecting Eq.~\eqref{eq:rel_mot} onto the directions $\hat{r}_1$ and $\hat{r}_2$, we obtain
\begin{align}
v_{1,\text{los}} -  v_{2,\text{los}} \cos\theta + v_{2,\perp} \sin\theta = \nonumber 
v_{\mathrm{rad}} ( \bar{r}_1 - \bar{r}_2 \cos\theta) + \tilde{v}_{\tan} \bar{r}_2 \sin\theta, \\
v_{2,\text{los}} - v_{1,\text{los}} \cos\theta - v_{1,\perp} \sin\theta = \nonumber v_{\mathrm{rad}} (\bar{r}_2 - \bar{r}_1 \cos\theta) - \tilde{v}_{\tan} \bar{r}_1 \sin\theta,
\end{align}

\noindent where $\bar{r}_1 = r_1/r_{12}$, $\bar{r}_2 = r_2/r_{12}$, and $\tilde{v}_{\tan} = v_{\tan} \sin i$ is the projection of the tangential velocity on the O12 plane. These expressions allow us to isolate $v_{\mathrm{rad}}$:

\begin{equation}
v_{\mathrm{rad}} = \frac{v_{2,\text{los}} -  v_{1,\text{los}} \cos\theta - v_{1,\perp}\sin\theta + \tilde{v}_{\tan} \bar{r}_2 \sin\theta}{\bar{r}_2 - \bar{r}_1 \cos\theta },
\label{eq:major_1}
\end{equation}

\begin{equation}
v_{\mathrm{rad}} = \frac{v_{1,\text{los}} - v_{2,\text{los}} \cos\theta + v_{2,\perp} \sin\theta - \tilde{v}_{\tan} \bar{r}_1 \sin\theta}{\bar{r}_1 - \bar{r}_2 \cos\theta }.
\label{eq:major_2}
\end{equation}

When either $v_{1,\perp}$ or $v_{2,\perp}$ and $v_{\tan}$ vanish, these equations simplify. The model assuming $v_{1,\perp} = v_{\tan} = 0$ is referred to as the major infall model. The two projection options (onto galaxy 1 or galaxy 2) are not symmetric. Assuming one galaxy has negligible perpendicular and tangential velocities, we can extract both $v_{\mathrm{rad}}$ and $\tilde{v}_{\tan}$:

\begin{align}
v_{\mathrm{rad}} = \ v_{1,\text{los}} \bar{r}_1 + v_{2,\text{los}} \bar{r}_2 
- \cos \theta ( v_{1,\text{los}} \bar{r}_2 + v_{2,\text{los}} \bar{r}_1 ) \nonumber \\
+ \sin \theta ( v_{2,\perp} \bar{r}_1 - v_{1,\perp} \bar{r}_2 ),
\label{eq:radial}
\end{align}

\begin{align}
\tilde{v}_{\tan} = \ v_{1,\perp} \bar{r}_1 + v_{2,\perp} \bar{r}_2 
- \cos \theta ( v_{1,\perp} \bar{r}_2 + v_{2,\perp} \bar{r}_1 ) \nonumber \\
+ \sin \theta ( v_{2,\text{los}} \bar{r}_1 - v_{1,\text{los}} \bar{r}_2 ).
\label{eq:tan}
\end{align}

In the minor infall model, where $v_{1,\perp} = v_{2,\perp} = 0$, only LoS velocities contribute. The radial velocity thus reduces to
\begin{equation}
v_{\mathrm{rad}} = \ v_{1,\text{los}} \bar{r}_1 + v_{2,\text{los}} \bar{r}_2 - \cos \theta \left( v_{1,\text{los}} \bar{r}_2 + v_{2,\text{los}} \bar{r}_1 \right).
\label{eq:minor_infall_append}
\end{equation}

We propose a new model where both galaxies move toward their common CoM. Assuming the CoM has only a LoS component, the velocity is written as

\begin{equation}
\vec{v}_{\text{c}} = v_{\text{c}} \hat{r}_{c}, \quad \vec{v}_{\text{c},\perp} = 0.
\label{eq:com_assum}
\end{equation}

The position and velocity of the CoM are given by

\begin{equation}
\vec{r}_{c} =  \bar{m}_1 \vec{r}_1 + \bar{m}_2 \vec{r}_2, \quad  \vec{v}_{c} =  \bar{m}_1 \vec{v}_1 + \bar{m}_2 \vec{v}_2,
\end{equation}where $\bar{m}_1 = \frac{m_1}{m_1 + m_2}$ and $\bar{m}_2 = \frac{m_2}{m_1 + m_2} = 1 - \bar{m}_1$. The relative velocities with respect to the CoM are $\bar{m}_2 v_{\mathrm{rad}}$ and $- \bar{m}_1 v_{\mathrm{rad}}$, giving

\begin{equation}
v_c \hat{r}_{c} + \bar{m}_2 v_{\mathrm{rad}} \hat{r}_{21} = \vec{v}_1, \quad
v_c \hat{r}_{c} - \bar{m}_1 v_{\mathrm{rad}} \hat{r}_{21} = \vec{v}_2.
\end{equation}

Projecting these onto $\hat{r}_1$ and $\hat{r}_2$ gives
\begin{equation*}
v_c (\bar{m}_1 \bar{r}_1+\bar{m}_2 \bar{r}_2 \cos\theta)+\bar{m}_2 v_\mathrm{rad} (\bar{r}_2 \cos\theta-\bar{r}_1) = v_{1,\text{los}},
\end{equation*}

\begin{equation}
v_c (\bar{m}_1 \bar{r}_1 \cos\theta + \bar{m}_2 \bar{r}_2)- \bar{m}_1 v_\mathrm{rad} (\bar{r}_2 - \bar{r}_1 \cos\theta ) = v_{2,\text{los}},
\end{equation}

\noindent which can be solved for $v_{\mathrm{rad}}$:

\begin{equation}
v_\mathrm{rad} = \frac{\bar{m}_1 \bar{r}_1 v_{2,\text{los}} -\bar{m}_2 \bar{r}_2 v_{1,\text{los}} - \cos\theta \left( \bar{m}_1 \bar{r}_1 v_{1,\text{los}} - \bar{m}_2 \bar{r}_2 v_{2,\text{los}}\right)}{\bar{r}_1 \bar{r}_2 \Delta \bar{m} + \cos\theta \left(\bar{m}_1 \bar{r}_1^2-\bar{m}_2 \bar{r}_2 ^2\right)},
\label{eq:radial_com_major_appen}
\end{equation}with $\Delta \bar{m} = \bar{m}_1 - \bar{m}_2$. We call this the PCM. For $\bar{m}_1 = 1$, this reduces to the major infall projection on galaxy 1, and for $\bar{m}_2 = 1$ on galaxy 2. This projection avoids a dependence on the unobserved perpendicular velocities. The model relies on knowledge of the mass ratio, which must be inferred or marginalized over using priors. The following identities support the analysis.

The unit vector $\hat{\boldsymbol{r}}_{21}$ from galaxy 1 to 2 is

\begin{equation}
\hat{\boldsymbol{r}}_{21} = \frac{\boldsymbol{r}_{2}- \boldsymbol{r}_{1}}{r_{21}} = \overline{r}_2 \hat{\boldsymbol{r}}_{2} - \overline{r}_1 \hat{\boldsymbol{r}}_{1}.
\label{eq:ur2}
\end{equation}The directions $\hat{\boldsymbol{r}}_i$ and $\hat{\boldsymbol{r}}_{\perp i}$ are orthogonal:

\begin{equation}
\hat{\boldsymbol{r}}_i \cdot \hat{\boldsymbol{r}}_{\perp i} = 0, \quad i =1,2.
\label{eq:ur3}
\end{equation}

The cosine of the angle $\theta$ between either pair of LoS or perpendicular directions is
\begin{equation}
\hat{\boldsymbol{r}}_1 \cdot \hat{\boldsymbol{r}}_2 = \hat{\boldsymbol{r}}_{\perp 1} \cdot \hat{\boldsymbol{r}}_{\perp 2} = \cos\theta.
\label{eq:ur4}
\end{equation}Similarly, the sine relations read
\begin{equation}
\hat{\boldsymbol{r}}_1 \cdot \hat{\boldsymbol{r}}_{\perp 2} = - \hat{\boldsymbol{r}}_2 \cdot \hat{\boldsymbol{r}}_{\perp 1} = \sin\theta.
\end{equation}Finally, the projections of $\hat{\boldsymbol{r}}_{21}$ along the LoS and perpendicular directions yield

\begin{equation}
\hat{\boldsymbol{r}}_1 \cdot \hat{\boldsymbol{r}}_{21} = \overline{r}_2 \cos \theta - \overline{r}_1, \quad
\hat{\boldsymbol{r}}_2 \cdot \hat{\boldsymbol{r}}_{21} = \overline{r}_2 - \overline{r}_1 \cos \theta,
\label{eq:ur10}
\end{equation}

\begin{equation}
\hat{\boldsymbol{r}}_{\perp 1} \cdot \hat{\boldsymbol{r}}_{21} = - \overline{r}_2 \sin \theta, \quad
\hat{\boldsymbol{r}}_{\perp 2} \cdot \hat{\boldsymbol{r}}_{21} = - \overline{r}_1 \sin \theta.
\end{equation}

\end{document}